# Performance Evaluation of Fuzzy Integrated Firewall Model for Hybrid Cloud based on Packet Utilization

Ziaur Rahman*, Asma Islam Swapna†, Habibur Rahman Habib‡ and Akramuzzaman Shaoun‡
*Department of Information and Communication Technology
Mawlana Bhashani Science and Technology University, Santosh, Tangail 1902
*Email: zia@iut-dhaka.edu
† Email: asma0swapna@gmail.com
‡ Email: mdhabibur.r.bd@ieee.org
§ Email: akramuzzamans@gmail.com

*Abstract*—Cloud computing is one of the highly flexible, confidential and easily accessible medium of platforms and provides powerful service for sharing information over the Internet. Cloud security has become an emerging issue as network manager eventually encounter its data protection, vulnerability during information exchange on the cloud system. We can protect our data from unwanted access on a hybrid cloud through controlling the respective firewall of the network. But, the firewall has already proved its weakness as it is unable to ensure multi-layered, secured accessibility of the cloud network. Efficient packet uti- lization sometimes causes high response time in accessing hybrid cloud. In this paper, a Cloud Model with Hybrid functionality and a secure Fuzzy Integrated Firewall for that Hybrid Cloud is proposed and thereby evaluated for the performance in traffic response. Experimental result illustrated that having a fuzzified firewall gives high point-to-point packet utilization  decreasing the response time than a conventional firewall. Results from this research work will highly be implemented in transplanting artificial intelligence in future Internet of Things (IoT).

## I. INTRODUCTION

Cloud is the way of reducing cost and effort for storage of large and big data. Cloud computing has a broader range of emerging applications, ranging from health care, government to financial markets. Organizations seek most effective cloud definition for its business purpose and profit matching their business requirements. Security and data access limits are the most prominent among them [1],[2].

Cloud can be of three types- Public, Private and Hybrid with respect to the organization needs and adaptions. Hybrid cloud is nothing but a mixed use of Public and Private Cloud network [2]. In order to get flexible data access from the network a well-structured Intrusion Prevention System (IPS) is a critical issue for a Hybrid Cloud [3]. These can be of two types- Network-based IPS and Host-based IPS. They both recognize the network attacks in two ways- known attack detection method and anomalous access detection methods. However, the mechanism for the attack detection needs to be perfect for better output [4],[5]. In this paper, the fuzzy logic is implemented as the attack detection logic of the firewall based IPS in the server side of the network in Hybrid Cloud.

Over the last 20 years or more firewall has changed a very little. Based on the previous researches there have a lot of studies went on over hybrid cloud efficiency in [6]. Furthermore, the firewall performance improvement over cloud network has been an issue for quite a long time as discussed in [7]. However, for the better performance evaluation and improvement of the firewall there is a need for an effective controller where the industry is larger. Fuzzy controller poses much stronger logical passage for the incoming and outgoing packets through the cloud [7].

Using member Function with Gaussian Mechanism input and taking the center of gravity as output in the model of Fuzzy controller, a Hybrid Cloud server side firewall is proposed in this paper. This paper specifically focuses on one aspect of that process a security model for hybrid cloud based system and internet firewall when it has no firewall, regular firewall and the implementation aspects of an automated fuzzy based firewall for web traffic block.

The remaining of the paper is organized as follows: Section II describes research background with relevance of  the proposed approach. Section III demonstrates modeling of Secured Cloud network and Fuzzy controller simulation setup. Evaluation and environment testing is described in section IV. Section V describes threat to the evaluation and limitation. Finally, the paper is concluded with future plan in section VI.

## II. BACKGROUND AND RELATED WORK

The concept of evaluating the performance of cloud computing based on a secure firewall implementation's protocol is presented in [8] to enhance necessary security requirements of hybrid clouds that efficiently blocks unwanted web traffic under particular attributes. Researchers also developed an on- tology based approach to analyze Cloud service compatibility using fuzzy logic for composing cloud optimization [9]. The firewall controls the incoming and outgoing traffic of an enterprise that stands between the internal network and the world outside for packet filtering which employed on large public networks. In our work, we have presented end-to-end utilization, throughput and response time of web traffic that is affected by the firewall and fuzzy integrated firewall.

Proposed fuzzy security algorithm for Internet firewall that can respond (between 0.7 and 1 it belongs to high security) in [10] to security level according to each packet dynamics state and take respective actions accordingly. Fuzzy logic is used to measure the security level of arriving packets. The firewall is fuzzily adaptive and proactive, intelligent, remains secure and speed, provides high security and high performance.

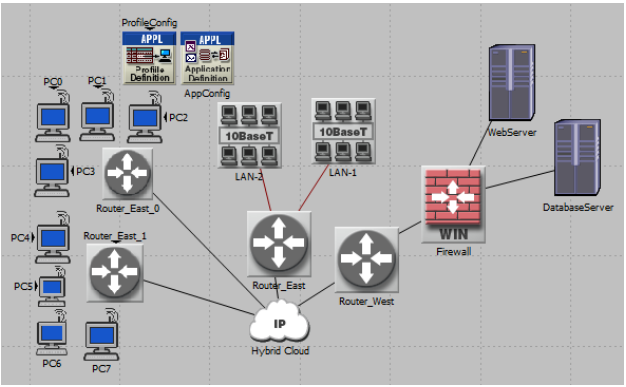

Fig. 1. Proposed Hybrid Cloud Architecture.

Based on the above, this paper evaluates the security issues, which are the biggest challenges for cloud computing. Therefore, the goal of this paper is to evaluate cloud computing with Riverbed modular simulator has been conducted in the cases of hybrid cloud computing regular and without firewall, fuzzy based firewall in a number of different scenarios to study the performance of the system in terms of delay, throughput and server traffic sent and received have been collected and enhanced when unauthorized web access is blocked by a fuzzy based firewall deployment model in a hybrid cloud [11].

III. PROPOSED APPROACH

Cloud is a metaphoric representation of Internet since the providers started using Virtual Private Network (VPN) [12]. Cloud is a distributed computing paradigm providing a wide distributed access to a more scalable virtualized hardware and software infrastructures to its huge range of users/clients [13]. Cloud computing has been a distributed computing paradigm nowadays [14]. Hybrid Cloud contains both private and pub- lic working data servers for its users. For the security of such private data server and the virtual infrastructure itself security measures vary with the needs. Firewall, VPN were vastly used previously [5]. The system is to be built here is based on the fuzzy logic implantation over the firewall. Fuzzy controller activity for system stability based on the rules for the output function to be produced works here for the firewall intelligence. Hence, the approach is to implement a system of cloud on hybrid sector which has the optimal access speed and as well the response time of data packets for data exchange through such cloud firewall. The logic inside the firewall will be based on the Fuzzy controller.

The system in consideration is the hybrid cloud. The cloud that has both public and private servers to perform commu- nication and data exchange. Figure 1 shows the provisional Hybrid Cloud to be implemented for secured fuzzified firewall with further performance evaluation.

The modeled cloud is the system where the proposed security framework has less limitation of scalability of the data packets. For the evaluation of packet passing security throughout the cloud servers both in private and public cloud there can be three states for the security appliance using firewall at the server sites that most matches the IPS on the network. These states are- On-state firewall, Off-state firewall and third is the firewall with special features [12]. This cloud model here holds the firewall with fuzzy controller to discard the unauthorized and anonymous web traffic, requests in the server both in the secured private cloud. However, the public sector of the cloud is open for all the traffic, hence left ahead of the outer dimension of firewall.

Very basic firewall communication for peer-to-peer traffic and perspective packet transfer works like the Figure 2 below. For all purposes traffic between sender ai and receiver bi the definition of the firewall includes traffic that gate, traverses the protected domain DA (ai , bi DA, 1). For the purpose of communication traffic between ai and bi that neither enters nor leaves the network does not belong to the firewalls technology.

However, if any traffic like in the Figure 2 one tries to enter the Hybrid Cloud that must lie and be passed through the logic gates set on the firewall for authentication and packet validation both for incoming and outgoing response packets [15]. The proposed firewall will carry the fuzzy agent and the member function will work based on the inputs to the controller and the output will be set as some security levels. The controller workflow diagram will work like the Figure 3.

Master Fuzzy Context (MFC) will be the predefined func- tions or set of rules in this cases that will impact on judging the optimal security level of the packets passing through the Fuzzy logic Controller. Based on the fuzzy agents (input packets), and the MFC the output will indicate the packet validation for entry and leave of the packets from or into the Hybrid Cloud.

The highest secured data destined to the private server, if requested to be accessed needs highest security level as the controller output and hence the packet filtration will be maximized. In this approach the proposed system architecture is illustrated in the Figure 4. Both firewall and web traffic with access is illustrated in the architecture.

Security threats and requirements for previous researches are analyzed for the enhancement of the existing system in this paper. Hence the proposed Cloud service security framework model of Hybrid Cloud Computing that includes Fuzzy related Firewall technology for implementation and ability to provide secured and enhanced performance for accessing the resource mobility in this model system approach.

IV. EVALUATION

To evaluate and implement the above proposed approach of Hybrid Cloud a fuzzy logic implementation is performed followed by the topological and network investigation using a simulator.

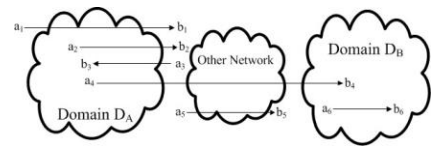

Fig. 2. Communication between Sender and Receiver through Firewall.

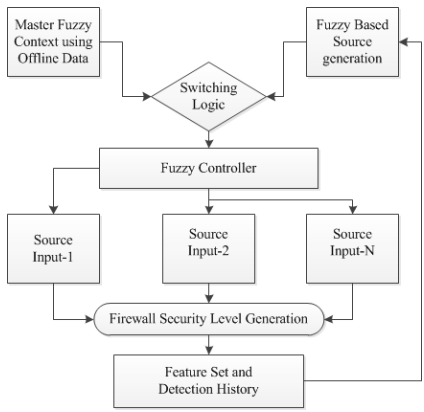

Fig. 3. Fuzzy Controlled Firewall Architecture.

## A. Environment Setup

For the mathematical method invocation on fuzzy control function with firewall rules for source and destination packet along with prospective security level MATLAB simulator is chosen for the sophisticated calculation and simulation of the logic controller [16]. The environmental setup process is performed for the Cloud Network topology using an effective simulator Riverbed for investigation specific implementation.

At the end of overall simulation the required data was collected for different graphical investigation regarding the Clouds performance. The performance criterion included the response time with fuzzified alongside the OFF and ON- state of cloud Firewall.

## B. Fuzzy Security Mechanism

In our proposed design, we have imphasised on the improvement of the security level of the Hybrid Cloud following to the rules generation. The fuzzified firewall will be better compared to the firewall and the no firewall. The security is measured both for the web server and the database server.

*1) Rule Generation:* The Master Fuzzy Context (MFC) will conduct as the member function for provisional fuzzy logic to be implemented using MATLAB in this paper. The rules to be set up in the input methods are formed with high and low secured data parameters and the output methods is the optimal security level. Hence, Table 1 holds the rules for the proposed cloud to be designed.

*2) Security Level:* According to the previously defined firewall controller rules in table 1 our simulation results are described in the following. Figure 5. It indicates the generated values with the security level as output. If the data packet is sent in small amount from the source and the destination get the data at the same rate, the data will be insecured. On the contrary if the data is sent either small or large amount from the sourse and the receiver received the data at high rate, it will be highly secured.

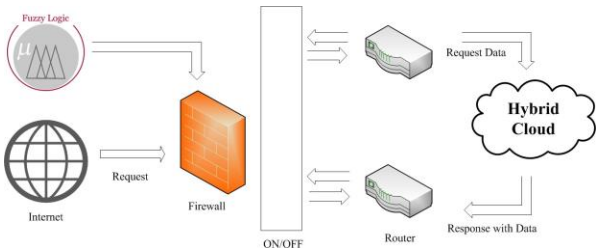

Fig. 4. System Architecture of the Secured Cloud.

TABLE I. SET OF FUZZY RULES TO INTEGRATE WITH FIREWALL.

| Source | Destination | Security |
|---|---|---|
| Low | Low | Insecure |
| Low | Medium | Low Security |
| Low | Medium-High | Medium Secured |
| Low | High | High Secured |
| Medium | Low-Medium | Medium Secured |
| Medium | Low | Insecure |
| High | High | High Secured |

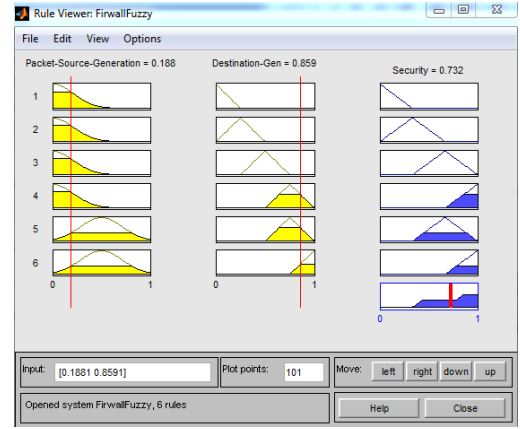

Fig. 5. Security Levels with Packet Source and Destination Observation.

## C. Result Analysis with Fuzzy Integration

The detailed simulation result analysis are from the discrete traffic analysis including page load time, traffic sent and receive rate on Hypertext Transfer Protocol (HTTP) request. Database query for the packet requests to the private severs and its responses were also collected and therefore represented in the graphs with these performance measurements.

In each figure the green line represents the values for no Firewall, blue represents the conventional firewall and red line indicates the designed Firewall with fuzzy logic. The results thar are measured on the basis of web server and database. The total time is calculated by the summing results of the response, send and receive time.

*1) Web Server:* In the Cloud Network HTTP request indicates the status of client and Cloud server connection status for information exchange [17]. For the proposed system HTTP request parameter is analyzed with its Page response time over the request of information, sent and received Traffic percentages over time.

Figure 6 shows there is a 10% decrease in amount of sent and receive traffic in bytes per second with fuzzified firewall than the sent and receive traffic rates with conventional firewall and 20% decreases than the cloud with no firewall. This indicates 10-20% high filtration in fuzzified Firewall after putting the fuzzy logic parameters in the designed cloud network. With overall response time for the previously re-quested HTTP packets indicate 12% less response time than the HTTP response time passing through the conventional firewall. However, the responses passing no firewall has 25% increased response time than that of fuzzified Firewall.

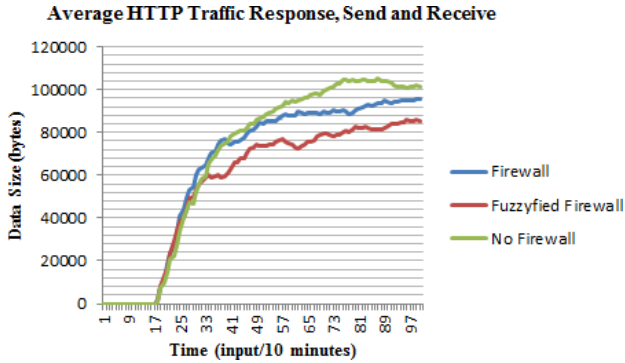

Fig. 6. Average Traffic Response, Send and Receive with Conventional Firewall, Fuzzified Firewall and No Firewall for HTTP.

*2) Database Server:* DB Query analysis for Database application is performed to investigate the database access parameters for the network in Riverbed. For the data access in the virtual cloud servers in the private subnetwork the simulation result is collected for database query packet requests, response. Hence, the analysis is performed for sent, received traffic along with perspective response time for the requests.

Figure 7 shows there is a 10-12% increase in amount of sent and receive traffic in bytes per second with fuzzified firewall than the sent traffic rates with conventional firewall and no firewall. This indicates 10-12% more authorized request packet passed through Fuzzified firewall while accessing the private cloud server.

The response time of the requested DB_Quary packets indicates 10% less response time than the query response time while hurling per the conventional firewall. The fuzzified firewall has 20% decrease response time than that the response time passing through the no firewall network.

Hence, though there are security levels up in the firewall,l the decreased response time and increased packet filtration indicates the systems stability and enhances Firewall performance along with the Hybrid Cloud with integrated fuzzy controller.

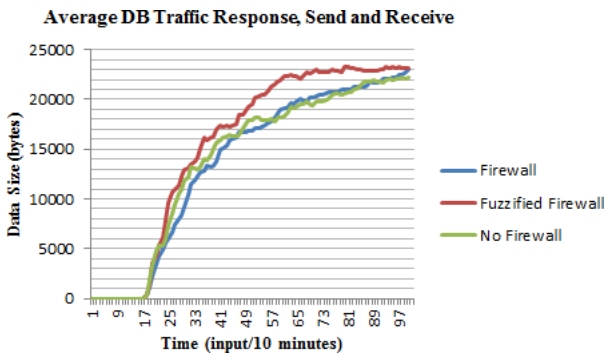

Fig. 7. Average Traffic Response, Send and Receive with Conventional Firewall, Fuzzified Firewall and No Firewall for DB Query.

## V. CONCLUSION AND FUTURE WORK

In this experimental research approach, fuzzy logic is integrated with firewall in a Hybrid Cloud. Firewall is the principal security aspect in this cloud. This paper evaluated the fuzzified firewall deployment model on a simulated hybrid cloud using a heavy load database application and a web application. Implementation and results as seen, are generated from OPNET simulator using three firewall scenarios: hybrid cloud with no firewall, hybrid cloud with conventional firewall and fuzzy integrated hybrid cloud firewall. The network is considered as a small cloud environment with 150 end nodes or less.

Experimental result showed having a conventional firewall in this hybrid cloud platform causes increased point-to-point utilization in web access and database access. However, firewall with fuzzy logic composed of security rules ended up showing 10% less response time compared to conventional firewall integration in the network model for accessing both web server applications and database applications. It would be interesting in future to apply this experimentation results in composing intellectual IoT network for home appliance or daily used connected devices. Further investigation in handling Big Data in such Hybrid Cloud platform will be another potential prospect of this experimentation.

## REFERENCES

[1] Y. Tang, P. P. Lee, J. Lui, and R. Perlman, "Secure overlay cloud storage with access control and assured deletion," *Dependable and Secure Computing, IEEE Transactions on*, vol. 9, no. 6, pp. 903–916, 2012.

[2] Q. Liu, C. Weng, M. Li, and Y. Luo, "An in-vm measuring framework for increasing virtual machine security in clouds," *Security & Privacy, IEEE*, vol. 8, no. 6, pp. 56–62, 2010.

[3] J. D. Burton, *Cisco security professional's guide to secure intrusion detection systems*. Syngress Publ., 2003.

[4] T. Sproull and J. Lockwood, "Wide-area hardware-accelerated intrusion prevention systems (whips)," in *Proceedings of the International Working Conference on Active Networking (IWAN)*, 2004, pp. 27–29.

[5] S. Dharmapurikar, P. Krishnamurthy, T. Sproull, and J. Lockwood, "Deep packet inspection using parallel bloom filters," in *High performance interconnects, 2003. proceedings. 11th symposium on*. IEEE, 2003, pp. 44–51.

[6] H. Song and J. W. Lockwood, "Efficient packet classification for network intrusion detection using fpga," in *Proceedings of the 2005 ACM/SIGDA 13th international symposium on Field-programmable gate arrays*. ACM, 2005, pp. 238–245.

[7] S. Jie, J. Yao, and C. Wu, "Cloud computing and its key techniques," in *Electronic and Mechanical Engineering and Information Technology (EMEIT), 2011 International Conference on*, vol. 1. IEEE, 2011, pp. 320–324.

[8] H. Kurdi, M. Enazi, and A. Al Faries, "Evaluating firewall models for hybrid clouds," in *Modelling Symposium (EMS), 2013 European*. IEEE, 2013, pp. 514–519.

[9] A. V. Dastjerdi and R. Buyya, "Compatibility-aware cloud service composition under fuzzy preferences of users," *IEEE Transactions on Cloud Computing*, vol. 2, no. 1, pp. 1–13, 2014.

[10] S.-U. Lar, X. Liao, A. ur Rehman, and M. Qinglu, "Proactive security mechanism and design for firewall," *Journal of Information Security*, vol. 2, no. 03, p. 122, 2011.

[11] *Riverbed Modular*, (accessed June 30, 2016). [Online]. Available: http://www.riverbed.com/sg/

[12] M. Sharma, H. Bansal, and A. K. Sharma, "Cloud computing: Different approach & security challenge," *International Journal of Soft Computing and Engineering (IJSCE)*, vol. 2, no. 1, pp. 421–424, 2012.

[13] J. Srinivas, K. V. S. Reddy, and A. M. QYSER, "Cloud computing basics," *International Journal of Advanced Research in Computer and Communication Engineering*, vol. 1, no. 5, 2012.

[14] Y. Jadeja and K. Modi, "Cloud computing-concepts, architecture and challenges," in *Computing, Electronics and Electrical Technologies (ICCEET), 2012 International Conference on*. IEEE, 2012, pp. 877–880.

[15] S.-H. Na, J.-Y. Park, and E.-N. Huh, "Personal cloud computing security framework," in *Services Computing Conference (APSCC), 2010 IEEE Asia-Pacific*. IEEE, 2010, pp. 671–675.

[16] *MATLAB Environment*, (accessed June 30, 2016). [Online]. Available: http://www.mathworks.com/products/matlab/

[17] S. Ray and A. De Sarkar, "Execution analysis of load balancing algorithms in cloud computing environment," *International Journal on Cloud Computing: Services and Architecture (IJCCSA)*, vol. 2, no. 5, pp. 1–13, 2012.

TABLE II. AUTHORS' BACKGROUND

| Name | Title | Research Field | Personal website |
|---|---|---|---|
| Ziaur Rahman | Phd candidate and Assistant Professor | Software Repository Mining | NA |
| Asma Islam Swapna | Undergrad Thesis Student | SDN, SDWN, WSN, Cloud Network, Network Performance, Traffic Engineering | www.asmaswapna.wordpress.com |
| Habibur Rahman | Undergrad Thesis Student | SDN, Cloud Network, Artificially Intelligent Network | NA |
| Akramuzzaman Shaoun | Undergrad Thesis Student | Future 5G network Performance, Wireless Network | NA |